\documentclass[preprint2]{aastex}

\newcommand{\psr}{\objectname{PSR B1451$-$68 }}

\shorttitle{PSR B1451$-$68 in X-rays}

\begin{document}


\title{Chandra observations of the old pulsar PSR B1451$-$68}

\author{B. Posselt}
\affil{Department of Astronomy \& Astrophysics, Pennsylvania State University, 525 Davey Lab,University Park, PA 16802, USA }
\email{posselt@psu.edu}
\author{G. G. Pavlov\altaffilmark{1}}
\affil{Department of Astronomy \& Astrophysics, Pennsylvania State University, 525 Davey Lab,University Park, PA 16802, USA}
\email{pavlov@astro.psu.edu}
\author{R. N. Manchester}
\affil{CSIRO Astronomy and Space Science, Australia Telescope National Facility, P.O. Box 76, Epping NSW 1710, Australia}
\author{O. Kargaltsev}
\affil{Department of Astronomy, University of Florida, 205 Bryant Space Center, Gainesville, FL 32611, USA}
\author{G. P. Garmire}
\affil{Department of Astronomy \& Astrophysics, Pennsylvania State University, 525 Davey Lab,University Park, PA 16802 , USA}
\altaffiltext{1}{St.-Petersburg State Polytechnic University, Polytekhnicheskaya ul. 29, 195251, Russia}

\begin{abstract}
We present 35\,ks \textit{Chandra} ACIS observations of the 42\,Myr old radio pulsar PSR B1451$-$68.
A point source is detected $0.32\arcsec \pm 0.73\arcsec$ from the expected radio pulsar position. It has $\sim 200$ counts in the 0.3-8\,keV energy range.
We identify this point source as the X-ray counterpart of the radio pulsar. 
\psr is located close to a 2MASS point source, for which we derive 7\% as the upper limit on the flux contribution to the measured pulsar X-ray flux. 
The pulsar spectrum can be described by either a power-law model with photon index $\Gamma=2.4^{+0.4}_{-0.3}$ and a unrealistically high absorbing column density  $N_{\rm H}=(2.5^{+1.2}_{-1.3}) \times 10^{21}$\,cm$^{-2}$, or by a combination of a $kT=0.35^{+0.12}_{-0.07}$ keV blackbody and a $\Gamma=1.4 \pm 0.5$ power-law component for $N_{\rm H}^{\rm DM}=2.6\times 10^{20}$\,cm$^{-2}$, estimated from the pulsar dispersion measure.
At the parallactic, Lutz-Kelker bias corrected distance of 480\,pc, the non-thermal X-ray luminosities in the 0.3-8\,keV energy band are either $L^{\rm nonth}_{\rm 0.3-8\,keV}=  (11.3 \pm 1.7) \times 10^{29}$\,erg\,s$^{-1}$ or  $L^{\rm nonth}_{\rm 0.3-8\,keV}=  (5.9^{+4.9}_{-5.0}) \times 10^{29}$\,erg\,s$^{-1}$, respectively. 
This corresponds to non-thermal X-ray efficiencies of either  $\eta^{\rm nonth}_{\rm 0.3-8\,keV}= L^{\rm nonth}_{\rm 0.3-8\,keV}/\dot{E} \sim 5 \times 10^{-3}$ or  $3 \times 10^{-3}$, respectively.
\end{abstract}




\section{Introduction}
Old ($>1$\,Myr), rotational-powered pulsars have lost a siginificant amount of their initial rotation energy. Their spin-down power, $\dot{E} \lesssim 10^{34}$\,erg\,s$^{-1}$, is usually several orders of magntiude lower than those of younger pulsars. The lower energy budget translates into fainter high-energy emission.  Only the closest of the old rotational-powered pulsars can be detected in X-rays.      
Since the bulk of the neutron star surface is too cold to be visible in X-rays \citep{yako2004},
one expects to see mainly the X-rays from the non-thermal magnetospheric emission
with a possible small thermal contribution from polar caps heated by infalling, accelerated particles \citep{Harding2001,Harding2002}.
So far, eight old pulsars have been detected at X-ray wavelengths (e.g., \citealt{Becker2004,Zavlin2004,Tep2005,Zhang2005,Becker2006, Kargaltsev2006,Misanovic2008,Hui2008,Pavlov2009}), several of them having only a few dozen counts. Most of the X-ray spectra can be described by relatively soft power laws with photon indices $2 \lesssim \Gamma \lesssim 4$. Some of these spectra can be fitted with blackbody (BB) models or by a combination of BB and power law (PL) components. The BB models usually indicate a rather small emission area compared to the conventional polar cap sizes whose radii are calculated  as $R_{PC}=[2\pi R_{\rm NS}^3/(cP)]^{1/2}$, assuming a magnetic dipole field \citep{Harding2001,Harding2002}. 
Here, $R_{\rm NS}$ is the radius of the neutron star, $P$ is the pulsar period and $c$ is the speed of light.  
There has been a debate as to how significant the thermal contributions are (e.g., \citealt{Becker2004,Zavlin2004,Misanovic2008,Hui2008}).
Better count statistics, the consideration of new information from other wavelengths as well as a larger object sample are required to understand the quantitative contributions of magnetospheric and thermal components.\\ 

PSR B1451$-$68 is an old, nearby pulsar, previously unexplored in X-rays.
The $0.263$\,s radio pulsar has a characteristic spin-down age $\tau = P (2\dot{P})^{-1} = 4.25 \times 10^7$\,yr, a dipole magnetic field strength of $1.63 \times 10^{11}$\,G, and a spin-down power $\dot{E} =2.1 \times 10^{32}$\,erg\,s$^{-1}$.
The proper motion components of \psr were measured by \citet{Bailes1990}: $\mu_{\alpha}= -39.5 \pm 0.7$\,mas\,yr$^{-1}$ and $\mu_{\delta}= -12.3 \pm 0.6$\,mas\,yr$^{-1}$. Unlike most other pulsars, the proper motion is directed toward the Galactic plane. \citet{Bailes1990} measured an annual parallax of $\pi = 2.2 \pm 0.3$\,mas and reported  a distance of $d=450 \pm 60$\,pc.
According to \citet{Verbiest2010} the distance corrected for the Lutz-Kelker bias is $d=480^{+80}_{-60}$\,pc.
The spin-down power of \psr is comparable to those of other nearby old pulsars which have been detected in X-rays. Aiming for constraints on the X-ray spectral properties of \psr, we report here on \textit{Chandra} observations of this pulsar.

\section{Observations and Data reduction}
\label{obs}
\psr was observed with the \textit{Chandra} X-ray observatory on 2010 May 16
using the ACIS-S detector in the VFAINT imaging mode. The nominal exposure time was 35.1\,ks.
The data reduction was done using CIAO (version 4.3).
We reprocessed the data in order to apply the VFAINT background cleaning.
We investigated event files with and without event position randomization using the Energy-Dependent Subpixel Event Repositioning (\texttt{EDSER}) algorithm by \citet{Li2004}, as well as the subpixel algorithm by \citet{Mori2001}. If not stated otherwise, we will discuss in the following the events with the pixel position randomization turned off.
No large background flares occured during the observation, therefore we filtered the reprocessed event file for good time intervals (GTIs) which have less the $5\sigma$ deviation from the overall lightcurve mean rate.
After the GTI filtering the actual exposure time was 34.7\,ks.
Due to the ACIS frame time of 3.2\,s, it is not possible to search for X-ray pulsations corresponding to the $P=0.263$\,s of PSR B1451--68.
The \texttt{wavdetect} task was applied to obtain an X-ray source list. 
A source was detected at the pulsar position on the back-illuminated chip S3.
Using the CIAO task \texttt{arfcorr} we estimate that 95\% of the source counts are included in a $r=2\arcsec$ aperture in an energy range from 0.3\,keV to 5\,keV. The aperture-corrected source count rate is  $0.0059 \pm 0.0004 $ cps.

Since there is a stellar neighbour with small angular separation ($\simeq 0\arcsec.7$) from the initial X-ray centroid position, we aimed for improving the astrometric solution of the event file.
As reference frames, we used the European Southern Observatory (ESO) New Technology Telescope (NTT) SUSI2 observations of \psr (PI: Bucciantini, obtained as part of their program 68.D-0249) and the 2MASS point source catalogue (PSC, \citealt{Skrutskie2006}). The details of the reprojection process are given in the Appendix~\ref{reproject}. 
Our derived overall \emph{absolute} astrometric uncertainty of the reprojected  \textit{Chandra} event file is $3 \sigma^{\rm AM}_{\rm abs}=0.73\arcsec$, and the \emph{relative} astrometric uncertainty of \textit{Chandra} with respect to the 2MASS PSC is $3 \sigma^{\rm AM}_{\rm rel}=0.66\arcsec$.

We obtained updated \texttt{wavdetect} positions for all X-ray sources on the S3 chip. The X-ray source at the pulsar location appears to be point like.
For this source, we also measured the centroid positions of the flux distribution in $r=2\arcsec$ and $r=0.6\arcsec$ apertures
for both, the event file with just the pixel position randomization removed and the one with the \texttt{EDSER} subpixel algorithm applied in addition. Statistical centroid position errors are calculated as the standard error of the mean in the pixel coordinates X and Y in the respective aperture and transformed to the world coordinate system.  
There are no statistically significant differences between the different positions listed in Table~\ref{table:positions}. 

\begin{table}[t]
\begin{center}
\caption{Chandra source positions in the reprojected event files.\label{table:positions}}
\begin{tabular}{lcc}
\tableline
method & R.A. & Dec. \\
 & $[ \circ ]$ & $[ \circ ]$ \\
\tableline
\multicolumn{3}{c}{event file without randomization} \\
\tableline
\texttt{wavdetect} & 224.00006(2) & -68.72757(1)\\
centroid, $r=0.6\arcsec$ & 224.00007(2) &  -68.72757(1) \\
centroid, $r=2\arcsec$   & 224.00002(2) &  -68.72759(1) \\
\tableline
\multicolumn{3}{c}{event file with \texttt{EDEF} subpixel algorithm employed} \\
\tableline
centroid, $r=0.6\arcsec$ &  224.00007(2) & -68.72757(1) \\
centroid, $r=2\arcsec$   &  224.00002(2) & -68.72758(1)\\
\end{tabular}
\tablecomments{Count-weighted variances of the centroid positions\\ are listed in brackets and refer to the last respective digits.\\ These errors do not include astrometric uncertainties.}
\end{center}
\end{table}

We extracted a source spectrum within $2\arcsec$ of the pulsar centroid position. 
A background spectrum was extracted from an annulus of $5\arcsec$ to $10\arcsec$ around the same position. 
The extracted source spectrum counts have been combined to energy groups of 15 counts each.
Applying XSPEC (version 12.6.0) to analyze the spectrum, we used $\chi^2$-fitting with standard weighting, the \texttt{tbabs} model for the X-ray absorption and abundance tables by \citet{Wilms2000} as well as photoelectric absorption cross sections from \citet{Balu1992} and \citet{Yan1998}. 
Different spectral models were checked for correspondence with the energy distribution of the $\sim 200$ source counts within an energy range of 0.3\,keV to 8\,keV.
The fit results are presented in Table~\ref{table:fits}. 
We checked whether the VFAINT background cleaning influenced the spectral properties of the source. Without this correction there are 2 more source counts (1\%), and the corresponding spectral model fit results are within $1\sigma$ the same as listed.

\section{Results and Discussion}
\subsection{Position and possible source blending}
\label{position}
In order to determine the pulsar coordinates, we combined the Molonglo data used by  \citet{Siegman1993} and Parkes data from 1991 to 2006.
First, we obtained the dispersion measure, ${\rm DM}=8.557\pm 0.014 (2\sigma)$\,cm$^{-3}$\,pc, using the Parkes data which have 400, 600 and 1400 MHz times of arrivals. This value improves the accuracy of the previously reported ${\rm DM}=8.6\pm 0.2 (2\sigma)$\,cm$^{-3}$\,pc by \citet{Aless1993}.
Fixing the proper motion at the interferometric values of \citet{Bailes1990}, and using the Molonglo as well as the Parkes data, we derive the position of \psr as RA(J2000)= 14:56:00.071(6); Dec(J2000)= $-$68:43:39.25(5) at epoch MJD 50135.
The quoted uncertainties are twice those given by the least-square solution.
The expected pulsar coordinates at the epoch of the \textit{Chandra} observations, MJD 55332.7, are: 
${\rm RA}_{\rm PSR}$ (J2000)= 14:55:59.968(8); 
$\rm{Dec}_{\rm PSR}$ (J2000)= $-$68:43:39.43(6). 
The errors reflect $2\sigma$ uncertainties including those due to the proper motion uncertainties.  

The measured X-ray source position slightly depends on the method applied, see Table~\ref{table:positions}. We will use in the following the \texttt{wavdetect} position in the reprojected event file without pixel position randomization. 
The measured X-ray source position is then 
${\rm RA}_{\rm CXO}=$ 14:56:0.015(7) and ${\rm Dec}_{\rm CXO}=-68$:43:39.25(4).
The errors reflect $2\sigma$ variances of the \texttt{wavdetect} position.

Close to this position there is a 2MASS point source which has also been detected in the ESO SUSI2 WB655 image as well as other optical surveys.
The listed J2000 position of 2MASS\,14560002$-$6843400 is 
${\rm RA}_{\rm 2MASS}$= 14:56:00.028; ${\rm Dec}_{\rm 2MASS}$= $-$68:43:40.02 \citep{Cutri2003}. 
2MASS\,14560002$-$6843400 is also listed in the USNO B1 catalogue \citep{Monet2003}.  \citet{Roeser2010} determined the apparent proper motion of this source to be $-0.6 \pm 8.2$\,mas yr$^{-1}$ in right ascension and $-1.9 \pm 8.2$\,mas yr$^{-1}$ in declination. Thus, we can neglect  movement of the 2MASS source, and the calculated separation between the expected pulsar position and the 2MASS point source is 0.68\arcsec\ .

The X-ray source position is offset by $0.26\arcsec \pm 0.06\arcsec$ in right ascension and $0.17\arcsec \pm 0.06\arcsec$ in declination from the expected pulsar position. Here, the errors reflect only the $3\sigma$ \texttt{wavdetect} positional variances. In addition, systematic astrometric errors apply. Our {\emph{absolute}} astrometric accuracy is better than $3 \sigma^{AM}_{abs}=0.73\arcsec$ (Section~\ref{obs} and Appendix~\ref{reproject}). Thus, the \textit{Chandra} X-ray position is coincident with the expected pulsar position within astrometric uncertainties.\footnote[1]{The separation between the position of 2MASS\,14560002$-$6843400 and the optical source detected in the ESO SUSI2 image is only 46\,mas, underlining the tight relation between the SUSI2 and 2MASS (and subsequently the \textit{Chandra}) astrometry.}

The X-ray source position is offset by\\ $0.07\arcsec \pm 0.06 \arcsec$ in right ascension and $0.76\arcsec \pm 0.06 \arcsec$  in declination  from the catalog position of 2MASS\,14560002$-$6843400.
To discuss the position of this 2MASS source, only the $3 \sigma^{AM}_{rel}=0.66\arcsec$  {\emph{relative}} astrometric precision between \textit{Chandra} and 2MASS has to be considered (Section~\ref{obs} and Appendix~\ref{reproject}). 
Considering the X-ray statistical (\texttt{wavdetect}) positional errors as well as the {\emph{relative}} astrometric precision between \textit{Chandra} and 2MASS, it appears unlikely that the \textit{Chandra} X-ray point source with 0.77\arcsec\ separation is the counterpart of the 2MASS point source.
We discuss spectral constraints on the nature of 2MASS\,14560002$-$6843400 in the Appendix~\ref{spec2}.
To estimate a limit of the possible X-ray flux contribution of 2MASS\,14560002-6843400 to the measured X-ray emission of the pulsar we use different approaches, which are discussed in detail in the Appendix~\ref{blending}.
In short, we measure flux percentages in small apertures centered on the X-ray source and the 2MASS source positions, deconvolve the image applying the \textit{Chandra} Ray Tracer (ChaRT), MARX and the CIAO task \texttt{arestore}, and investigate the count energy distribution of the X-ray source.  
Considering (i) the flux percentage measurements in the reprojected event file with or without subpixel algorithm applied; (ii) the comparison of the deconvolved image with an deconvolved \emph{simulated} double source image; (iii) the location of the PSF asymmetry region; and (iv) the hardness of the detected counts towards the 2MASS star; we conclude that any contribution of  2MASS\,14560002-6843400 to the measured X-ray flux of the pulsar must be $<7$\%.

\subsection{The X-ray spectrum of the pulsar}
\label{thepulsar}
As described in Sect.~\ref{obs}, we used XSPEC to check different spectral models for the extracted X-ray source spectrum.
As discussed above, at least 93\% of the X-ray source flux is attributed to the pulsar. 
We employ in the following the counts from the $r=2\arcsec$ extraction region for the pulsar spectrum (198 counts in an energy range from 0.3\,keV to 5\,keV) and neglect the possible (small) contribution of the 2MASS star.  
The fit results are presented in Table~\ref{table:fits}. 
The derived absorbed X-ray fluxes for the different spectral models range from $2.4 \times 10^{-14}$\,erg\,cm$^2$\,s$^{-1}$ to $2.9 \times 10^{-14}$\,erg\,cm$^2$\,s$^{-1}$ for energies between 0.3\,keV and 5\,keV.
Allowing all parameters to vary, we tested simple single-component models like a BB or a PL spectrum. An absorbed, $N_{\rm H}=2.5^{+1.2}_{-1.3} \times 10^{21}$\,cm$^{-2}$, 
 PL model with photon index $\Gamma =2.4^{+0.4}_{-0.3}$ fits the spectrum well, while the obtained BB fit is statistically unacceptable. 
In addition, we checked whether the X-ray spectrum can be described by an optically thin thermal plasma model as one could expect from a stellar corona. 
The \texttt{apec} plasma model fit is acceptable. It is slightly worse than the PL, but better than the BB fit. We discuss the apec parameters with respect to 2MASS\,14560002$-$6843400 in more detail in Appendix~\ref{spec2}. For the pulsar counterpart of the X-ray source, the thin plasma model is physically irrelevant and not further discussed.\\

\begin{deluxetable}{lcccccccc}
\tabletypesize{\footnotesize}
\rotate
\tablecaption{XSPEC fit results\label{table:fits}}
\tablewidth{0pt}
\tablehead{
\colhead{Model} & \colhead{$N_{\rm H}$} & \colhead{$\Gamma$} & \colhead{PL norm at 1\,keV} & \colhead{$kT$} & \colhead{$R_{BB}$} & \colhead{EM} & \colhead{red. $\chi^2$/d.o.f.} & \colhead{$F_{unabs}$}\\
\colhead{ } & \colhead{$[10^{20}$\,cm$^{-2}]$} & \colhead{ } & \colhead{$[10^{-6}$photons\,keV$^{-1}$\,cm$^{-2}$\,s$^{-1}]$}  & \colhead{[keV]} & \colhead{[m]} &\colhead{$[10^{38}$\,cm$^{-3}]$} & \colhead{ } & \colhead{$[10^{-14}$\,erg\,cm$^2$\,s$^{-1}]$} 
}
\startdata
PL & $25^{+12}_{-13}$ & $2.4^{+0.4}_{-0.3}$ & $13^{+6}_{-4}$ & \ldots & \ldots & \ldots & 0.4/10 & $6.0^{+2.1}_{-1.4}$\\
BB & $\leq 12$ & \ldots & \ldots & $0.42^{+0.06}_{-0.05}$ & $13.2^{+11.2}_{-7.2}$ & \ldots & 1.4/10 & $2.4^{+0.4}_{-0.3}$\\
BB+PL & $17^{+26}_{-17}$ & $2.2 \pm 1.0$ & $8.2^{+10}_{-8.2}$ & $0.3 \pm 0.2$ & $9.9^{+38.2}_{-9.9}$ &  \ldots &0.5/8 & $3.8^{+5.2}_{-2.8}$ (PL) / $0.7^{+1.4}_{-0.7}$ (BB)\\
APEC & $6.5^{+7.9}_{-4.7}$ & \ldots & \ldots & $3.1^{+1.3}_{-0.8}$ &\ldots & $6.6 \pm 1.0$ & 0.8/10 &$3.8^{+0.7}_{-0.6}$\\
\tableline
 & frozen\\
\tableline
PL & 2.6 & $1.7 \pm 0.2$ & $6.6 \pm 0.9$ & \ldots & \ldots & \ldots & 1.6/11 & $4.1 \pm 0.6$\\
BB & 2.6 & \ldots & \ldots &  $0.40^{+0.06}_{-0.05}$ & $14.1^{+12.3}_{-8.3}$ & \ldots & 1.4/10 & $2.4^{+0.4}_{-0.3}$\\
BB+PL & 2.6 & $1.4 \pm 0.5$ & $2.7^{+2.1}_{-2.7}$ & $0.35^{+0.12}_{-0.07}$ & $13.8^{+24.2}_{-12.3}$ & \ldots & 0.5/9 & $2.1 \pm 1.1$ (PL) / $1.4^{+0.9}_{-0.7}$ (BB)\\ 

\enddata
\tablecomments{All errors indicate 90\% confidence intervals, the BB emitting area radius errors take the distance error into account. The parallactic distance, corrected for the Lutz-Kelker Bias, $D=480^{+80}_{-60}$\,pc, is used. The unabsorbed fluxes, $F_{unabs}$, are given for the energy range from 0.3\,keV to 8\,keV.}
\end{deluxetable}

\begin{figure}[t]
\begin{center}
\includegraphics[angle=-90,width=7.5cm,bb=60 -6 580 710,clip]{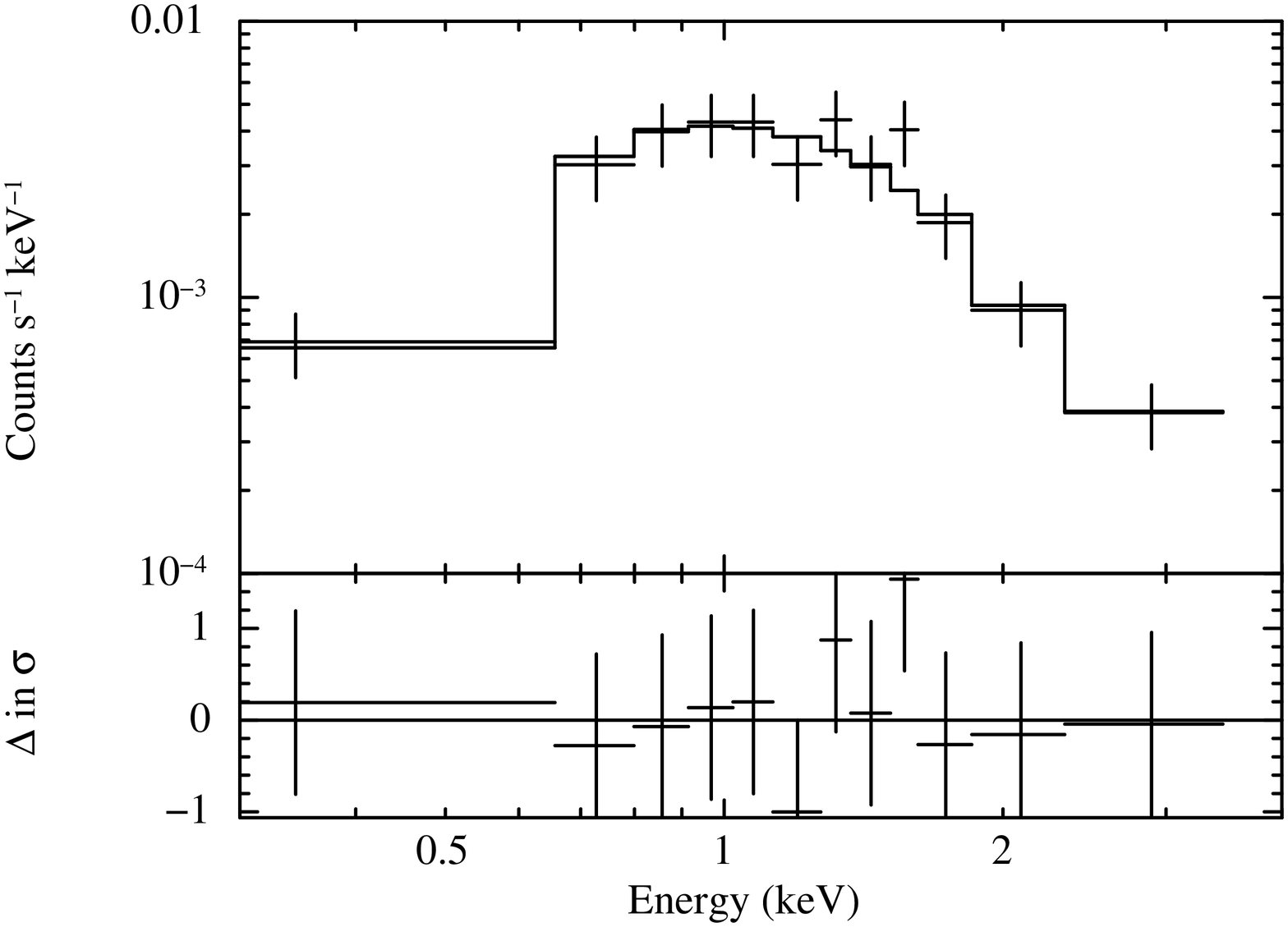}
\includegraphics[angle=-90,width=7.5cm,bb=79 20 595 701,clip]{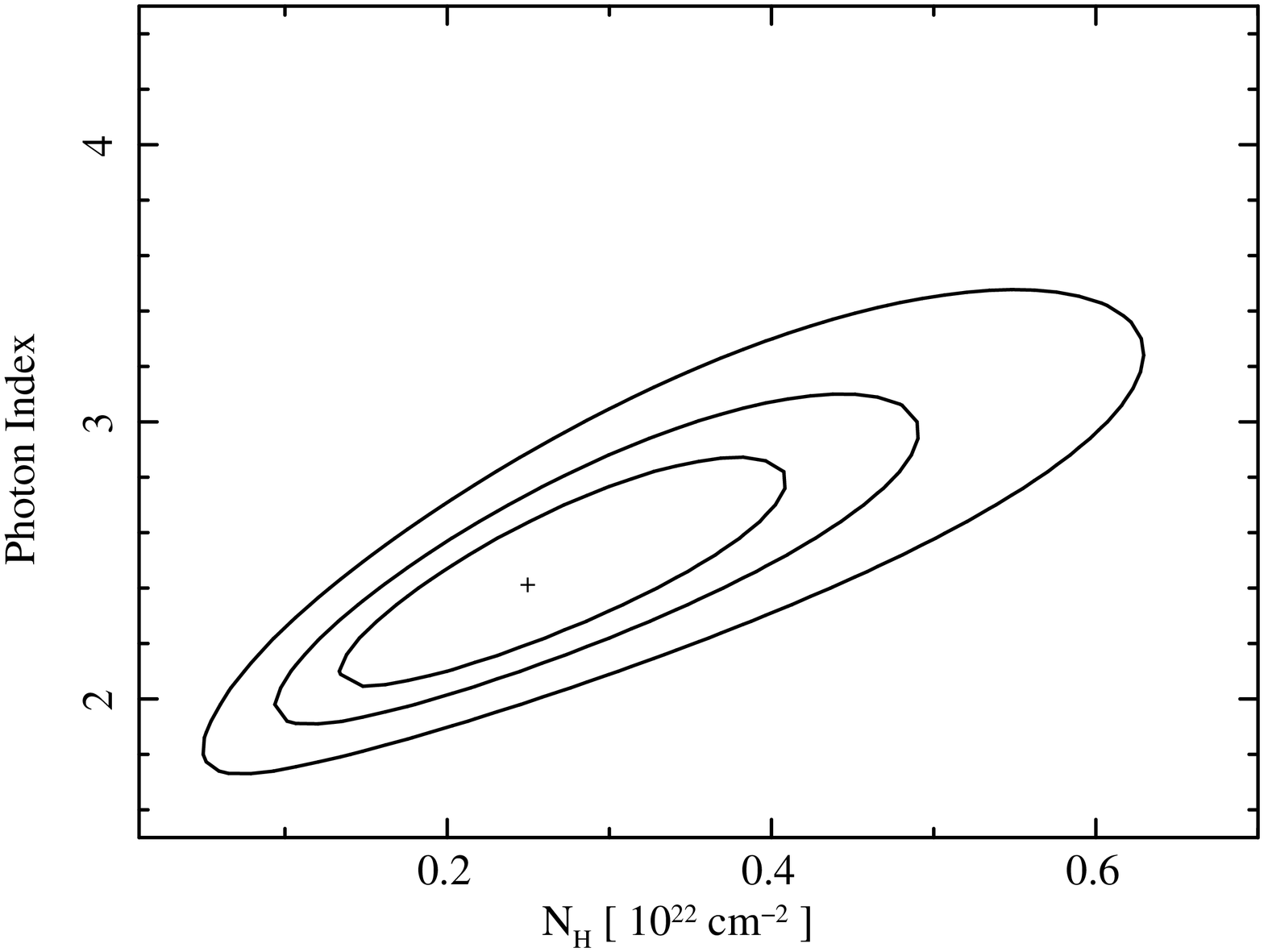}
\end{center}
\caption{X-ray spectrum of PSR B1451$-$68 and the best fit absorbed PL model if $N_{\rm H}$ is a free fit parameter; see text and Table~\ref{table:fits}. The lower panel of the upper plot shows the residuals in units of $\sigma$. The lower plot shows 68\%, 90\%, 99\% confidence contours in the $N_{\rm H}-\Gamma$ plane of the PL model for the pulsar spectral data.\label{powfit}}
\end{figure}

Among the X-ray spectral fits in Table~\ref{table:fits}, the PL fit and the BB+PL fit show the lowest $\chi^2$ values, but the uncertainties are very large for the two-component model. 
Both fits 
show suspiciously high best-fit values of the hydrogen column density, $N_{\rm H} \sim 2 \times 10^{21}$\,cm$^{-2}$. This value is close to the Galactic value for this line of sight ($l=313.87^\circ$, $b=-8.54^\circ$): the LAB Survey of Galactic neutral hydrogen reports $1.6\times 10^{21}$\,cm$^{-2}$ towards this direction \citep{Kalberla2005}, the \citet{Dickey1990} neutral hydrogen survey reports $2.1 \times 10^{21}$\,cm$^{-2}$.
The pulsar has a parallactic distance of only 450\,pc and a low DM value.
Assuming 10 H atoms for each $e^-$, we derive an expected $N_{\rm H}^{\rm DM}=2.6 \times 10^{20}$\,cm$^{-2}$ from the pulsar ${\rm DM}=8.557\pm 0.014$\,pc\,cm$^{-3}$ (Sect.~\ref{position}). 
Freezing $N_{\rm H}$ at $=2.6 \times 10^{20}$\,cm$^{-2}$, we obtain new X-ray spectral fits, presented in Table~\ref{table:fits}. The PL fit is now slightly worse than the BB fit, both are unacceptable. The combined BB+PL model fits the data best. According to the F-test, the combined model fits the data better than the PL-only fit with a probability of 99.4\%, and better than the blackbody-only fit with a probability of 99.1\%. 
\begin{figure}[t]
\includegraphics[angle=-90,width=7.7cm,bb=73 1 598 739,clip]{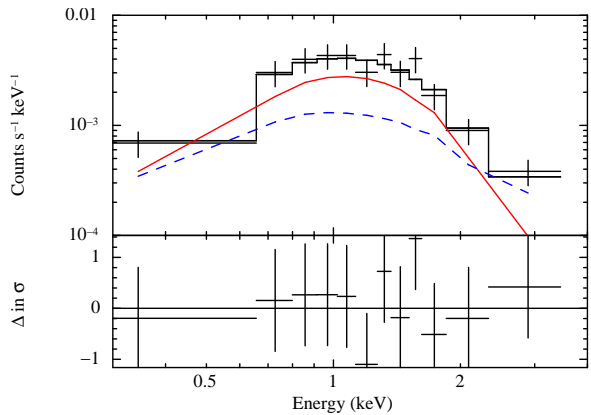}
\caption{X-ray spectrum of PSR B1451$-$68 and the best-fit absorbed BB+PL model in case of frozen $N_{\rm H}=2.6 \times 10^{20}$\,cm$^{-2}$.
The red, solid and blue, dashed lines visualize the BB model component and PL model component contributions, respectively.
The lower panel shows the residuals in units of $\sigma$. For more details see text and Table~\ref{table:fits}.}
\label{combfitnhfest}
\end{figure}
 
The inferred BB emitting area radius is of a similar size as in the case of the BB-only fit, albeit with larger error.
The  individual contributions to the unabsorbed fluxes in the energy range from 0.3\,keV to 8\,keV are $F^{\rm BB}_{X,unabs}=(1.4^{+0.9}_{-0.7}) \times 10^{-14}$\,erg\,cm$^2$\,s$^{-1}$ and $F^{\rm PL}_{X,unabs}=(2.1 \pm 1.1) \times 10^{-14}$\,erg\,cm$^2$\,s$^{-1}$  for the BB and the PL components, respectively.
Flux errors are large because the fitted component normalization values have high uncertainties due to the low number of counts.\\ 

We also checked neutron star (NSA) models \citep{Zavlin1996,Pavlov1995} with fixed parameters for neutron star mass and radius  (M$_{\rm NS}=1.4$\,M$_{\odot}$, R$_{\rm NS}=12$\,km). The fits are marginally acceptable, and the non-magnetisied NSA model fit achieves the lowest $\chi^2$. However, the fits are worse than the BB+PL fit. Allowing both the mass and radius to vary results in inplausibly low masses and radii.\\

The inferred X-ray spectral fit parameters are similar to those found for other old pulsars (e.g., \citealt{Pavlov2009,Kargaltsev2006,Becker2004}). From a statistical viewpoint alone we are unable to differentiate between a purely magnetospheric emission model and a combination of magnetospheric and thermal emission.
But the derived $N_{\rm H}$ in case of the PL-only fit is a factor 10 higher than one would expect from the dispersion measure. Even accounting for the fit parameter error and a possible uncertainty factor of 3 for the DM-based $N_{\rm H}$, these values do not overlap. Therefore, we regard the spectral fits, in particular the BB+PL fit, with fixed $N_{\rm H}$ more realistic.
In the combined model, the photon index of $\Gamma=1.4\pm 0.4$ is comparable to the typical values  for younger pulsars ($0.6<\Gamma < 2.1$, \citealt{Kargaltsev2008,Gotthelf2003}), but the inferred BB emission area radius, $13.8^{+24.2}_{-12.3}$\,m is rather small. The conventional polar cap (PC) radius is $R_{\rm PC} \simeq 280$\,m assuming a dipolar magnetic field for this pulsar and $R_{\rm NS}=10$\,km. 

Similar discrepancies have been found for BB components of other old pulsars \citep{Pavlov2009, Misanovic2008, Kargaltsev2006, Zhang2005}. 
Apart from geometrical projection effects, the following explanations have been discussed: 
only a small fraction of the PC is heated by inflowing particles at the footpoints of ``spark discharges'' created above the PC \citep{Zhang2005, Ruderman1975}; 
the PCs are covered by hydrogen or helium atmosphere resulting in effective temperatures of a factor 2 less and a radius a factor 3-10 larger than the BB fit \citep{Zavlin2004}; 
the nondipolar component of the magnetic field is much stronger than the dipolar component, causing partially screened acceleration regions and smaller PCs as consequences \citep{Gil2003}.\\

\subsection{The X-ray efficiency of old rotation-powered pulsars}
\begin{figure}[h]
\includegraphics[width=5.6cm, angle=90]{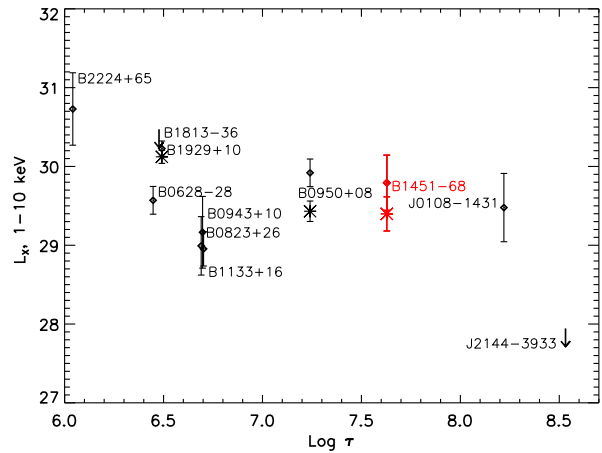}
\caption{X-ray luminosities and upper limits of eleven old pulsars versus their characteristic ages. In case of a sufficient number of counts, the diamonds and asterisks show the non-thermal and thermal X-ray luminosities, respectively; otherwise the luminosities were obtained from PL fits. The arrows mark upper limits derived from X-ray non-detections. This figure is an update of the one presented earlier by \citet{Kargaltsev2006} using new distances by \citet{Deller2009}, the Lutz-Kelker bias corrections by \citet{Verbiest2010}, and the upper limit result for PSR\,J2144-3933 by \citet{Tiengo2011}. Note that the large X-ray luminosity of the extremely fast moving  PSR B2224+65 may be  overestimated due to a suspected, unresolved shocked pulsar wind component (e.g., Johnson \& Wang 2010).}
\label{oldlumage}
\end{figure}

\begin{figure}[h]
\includegraphics[width=5.6cm, angle=90]{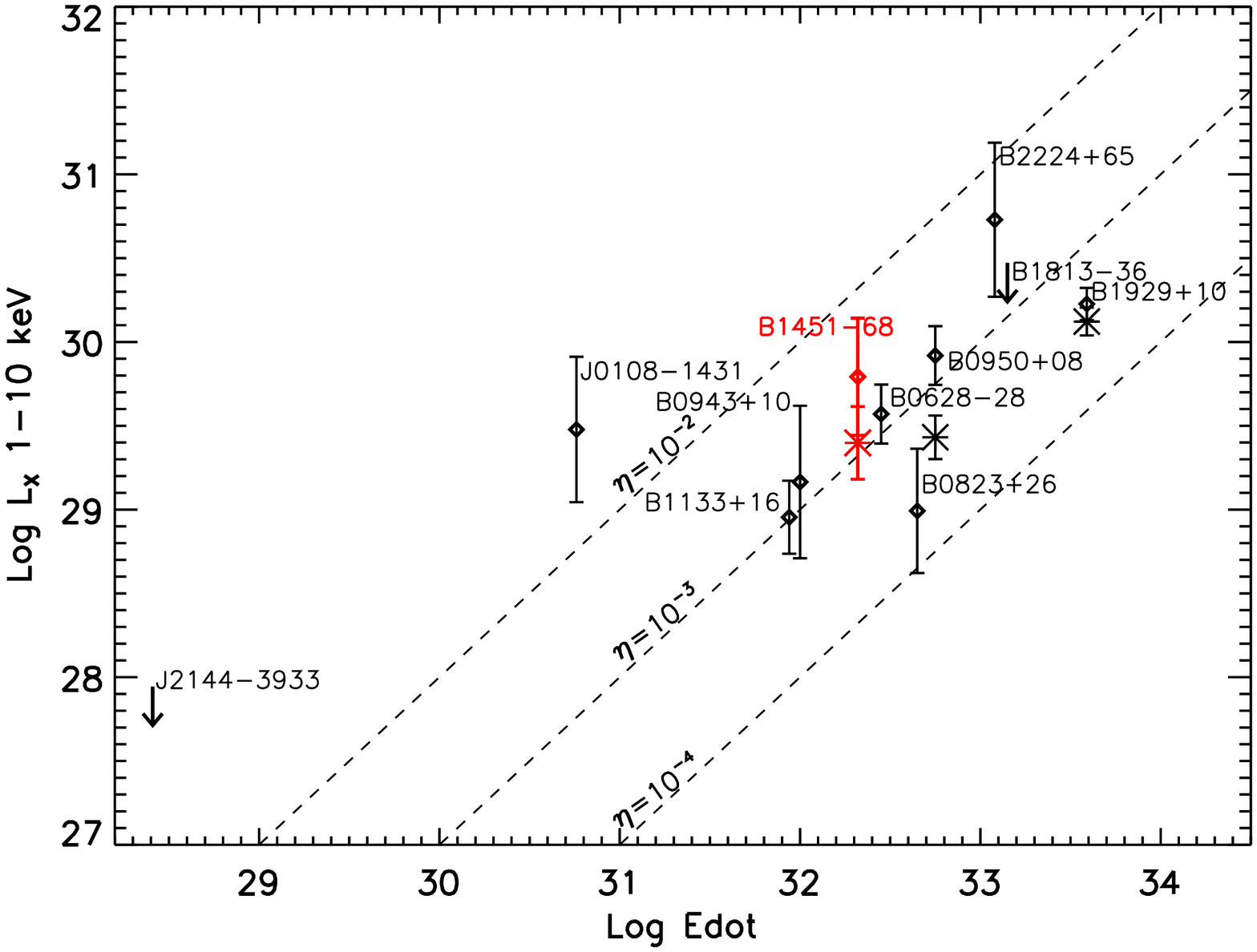}
\caption{X-ray luminosities and upper limits of eleven old pulsars versus their spin-down power. In case of a sufficient number of counts, the diamonds and asterisks show the non-thermal and thermal X-ray luminosities, respectively; otherwise the luminosities were obtained from PL fits. The arrows mark upper limits derived from X-ray non-detections.
This figure is an update of those presented earlier by \citet{Kargaltsev2006} and \citet{Pavlov2009} using new distances by \citet{Deller2009}, the Lutz-Kelker bias corrections by \citet{Verbiest2010}, and the upper limit result for PSR\,J2144-3933 by \citet{Tiengo2011}. Note that the large X-ray luminosity of the extremely fast moving  PSR B2224+65 may be  overestimated due to a suspected, unresolved shocked pulsar wind component (e.g., Johnson \& Wang 2010).}
\label{oldlumedot}
\end{figure}

\begin{figure*}[h]
\includegraphics[width=12cm, angle=90]{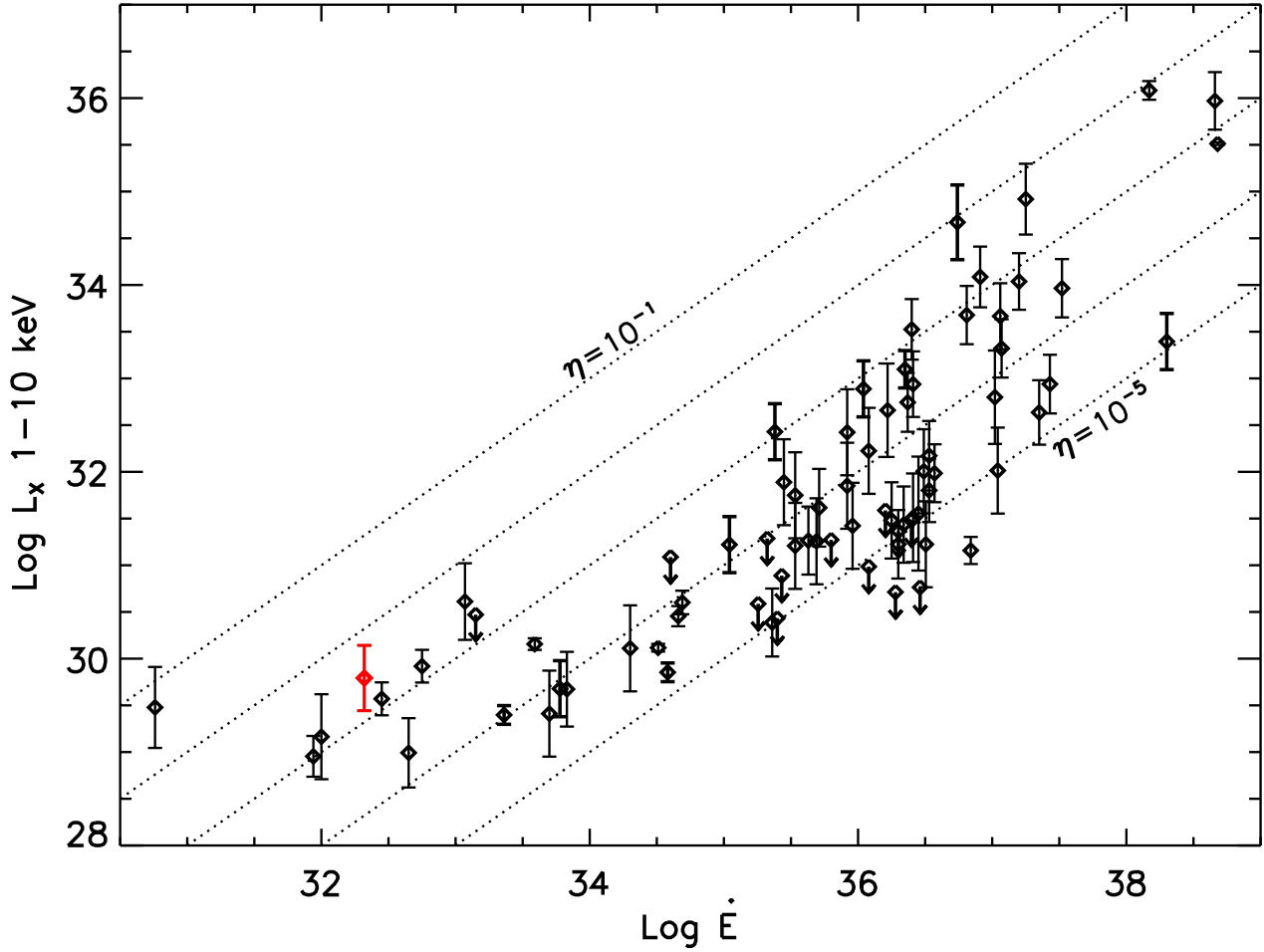}
\caption{Non-thermal X-ray pulsar luminosities and upper limits versus the pulsar spin-down power. 
This plot is an update of Figure 5 by \citet{Kargaltsev2008} including the old pulsars from the previous figure, recycled X-ray pulsars and new X-ray detected Fermi pulsars. The latter will be presented in detail in Kargaltsev et al. 2012
(submitted). As in the previous figure, parallactic distances were corrected for the Lutz-Kelker bias, where applicable.}
\label{oldlumedotall}
\end{figure*}

For PSR\,B1451--68, the luminosity of the non-thermal (PL) X-ray component in the 0.3-8\,keV energy band is $L^{\rm nonth}_X=4 \pi d^2 F^{\rm PL}_{X,unabs} = 5.9^{+4.9}_{-5.0} \times 10^{29}$\,erg\,s$^{-1}$, where the error is calculated from the 90\% confidence error of the unabsorbed flux and the $80$\,pc uncertainty of the Lutz-Kelker corrected distance.
Such a luminosity corresponds to a non-thermal X-ray efficiency $\eta_{\rm nonth}= L^{\rm nonth}_X/\dot{E} \sim 3 \times 10^{-3}$. 
The X-ray luminosities versus characteristic ages of old pulsars, including PSR\,B1451--68, are plotted in Figure~\ref{oldlumage}. There is no obvious characteristic age effect visible in this plot.
We also include \psr in the plot showing the X-ray luminosities versus spin-down energies of old pulsars (Figure~\ref{oldlumedot}). Distances to other old pulsars are updated according to \citet{Deller2009} and corrected for the Lutz-Kelker bias \citep{Verbiest2010}.   
The non-thermal X-ray efficiency of \psr is larger than those of young pulsars, most of which have $\eta_{\rm nonth}$ values smaller than $10^{-3}$, see Figure~\ref{oldlumedotall}.
The efficiency of PSR\,B1451--68 is, however, comparable to those of other old pulsars 
(Fig.~\ref{oldlumedot}). While it is possible that old pulsar spectra with too few counts may have an undetected thermal contribution (e.g., from heated polar caps), Figure~\ref{oldlumedot} shows only little influence of such a thermal component on the inferred non-thermal X-ray efficiencies in the energy range 1 to 10\,keV for the 3 pulsars with sufficiently large number of counts.
This supports the hypothesis that the conversion of spin-down power into X-ray emission becomes more efficient as pulsars get less powerful \citep{Zharikov2006,Kargaltsev2006}.
There is, however, an observational bias in favor of brighter X-ray sources which can lead to the detection of only the most efficient old neutron stars.

A look on the increasingly complete observational statistics of the \emph{closest} pulsars can help to minimize the effect of the observational bias in favor of brighter X-ray sources. Instead of the characteristic ages, which may be very different from the true ages, we prefer to use the directly measurable $\dot{E}$ as parameter to differentiate between powerful, young and less powerful, old pulsars.  
We choose $\dot{E}= 9 \times 10^{33}$\,erg\,s$^{-1}$ as a rough, admittedly arbitrary boundary between young and old pulsars. 
This boundary was motivated by pulsar statistics in the ATNF pulsar catalog\footnote[2]{http://www.atnf.csiro.au/research/pulsar/psrcat/} \citep{Manchester2005}, where most radio pulsars have characteristic ages $\tau \ge 1$\,Myr
for $\dot{E} < 9 \times 10^{33}$\,erg\,s$^{-1}$ while younger radio pulsars usually have higher $\dot{E}$.
The close $\gamma$-ray pulsar PSR J1741-2054 with $\tau=386$\,kyr, but modest  $\dot{E}=9.5 \times 10^{33}$\,erg\,s$^{-1}$ influenced our choice of the particular boundary value.

According to the ATNF pulsar catalog,  
PSR\,B1451--68 with its Lutz-Kelker corrected distance of $480^{+80}_{-60}$\,pc is the eleventh closest of the isolated radio pulsars with $\dot{E}< 9 \times 10^{33}$\,erg\,s$^{-1}$. 
Six of these eleven pulsars\footnote[3]{PSR B2224+65 with its potentially overestimated X-ray luminosity due to an unresolved shocked pulsar wind component is not among the eleven closest old pulsars.} are now detected in X-rays, all six have parallactic distances.
  Two of the X-ray detected pulsars,  PSR\,B1929+10 and PSR\,B0823+26, have X-ray efficiencies $10^{-4} < \eta_{\rm nonth} < 10^{-3}$, the other four X-ray detected pulsars have  $\eta_{\rm nonth} > 10^{-3}$. Thus, at least 36\% of all the closest old pulsars with  $\dot{E}< 9 \times 10^{33}$\,ergs\,s$^{-1}$ show  $\eta_{\rm nonth} > 10^{-3}$. 

For comparison, we consider now the 11 closest, isolated, younger pulsars with $\dot{E}> 9 \times 10^{33}$\,ergs\,s$^{-1}$. 
Nine of them are detected in X-rays, the two others were not probed deep enough.
Four of these nine detected pulsars have parallactic distances.  
Based on information of the ATNF pulsar catalog, X-ray flux investigations by \citet{Marelli2011,Camilo2009,Kargaltsev2008,Tep2007}, and distance corrections according to \citet{Verbiest2010} and \citet{Mignani2010}, we find that all nine X-ray detected younger pulsars have $\eta_{\rm nonth} < 10^{-3}$. Thus, at most 18\% of all these 11 pulsars with  $\dot{E}> 9 \times 10^{33}$\,ergs\,s$^{-1}$ could have  $\eta_{\rm nonth} > 10^{-3}$. 

Thus, looking only at the 11 closest sources in each case, it is suggestive, that at least one third of old pulsars have efficiencies $\eta > 0.001$, while none of the detected closest, younger pulsars has.   
Recently, \citet{Marelli2011} reported the following $L_X$-$\dot{E}$ relation for 29 Fermi pulsars with available distance estimates: 
$\log_{10} L_{X,29} = (1.11^{+0.21}_{-0.30}) + (1.04 \pm 0.09) \log_{10} \dot{E}_{34}$, which is in agreement with older estimates for X-ray pulsars by \citet{Kargaltsev2008} and \citet{Possenti2002}. From this, one would indeed expect $\eta < 10^{-3}$ for most X-ray pulsars, and the $>36$\% deviation of the closest, low $\dot{E}$ pulsars appear even more puzzling. 
Unfortunately, an insufficient number of probed objects, inhomogeneous depth of the X-ray observations, low signal-to-noise ratios in many cases, potentially unresolved compact pulsar wind nebulae, as well as large uncertainties of DM-based distances, prohibit currently any further conclusive statistics, for a larger sample of old pulsars in particular. 
Whether there is indeed a higher X-ray efficiency for pulsars with low $\dot{E}$ can only be checked by increasing the sample of such pulsars with sufficiently sensitive X-ray observations.

\section{Summary}
We investigated \textsl{Chandra} observations of PSR B1451$-$68.
Through various image analysis techniques we estimate the possible contribution from a nearby 2MASS star to the pulsar's X-ray flux to be less than 7\%.
No significant extended emission is seen. The pulsar has a soft X-ray spectrum similar to those of other old pulsars. Nominally, the spectrum is best fit with a power law, having a photon index of $\Gamma=2.4^{+0.4}_{-0.3}$. However, the inferred hydrogen column density, $N_{\rm H}$, is uncomfortably close to the Galactic value of HI in this direction. Fixing  $N_{\rm H}$ to the DM-derived $2.6 \times 10^{20}$\,cm$^{-2}$, a combination of a thermal and a non-thermal component
fits the data best. The inferred radius of a heated polar cap is small, a factor 20 less than expected conventionally.  
The nonthermal X-ray efficiency, $\sim 3 \times 10^{-3}$, is high in comparison to younger pulsars, but comparable to estimates for other old pulsars.

\medskip

\acknowledgments
We thank P. Broos, K. Getman and L. Townsley for enlighting discussions regarding Chandra data reduction. We also thank Elizabeth Galle and her colleagues from the CXC helpdesk for helpful suggestions regarding the reprojection of the Chandra event file.\\
Support for this work was provided by the ACIS  Instrument Team contract SV4-74018 (PI: G. Garmire) issued by the Chandra X-ray Observatory Center, which is operated by the Smithsonian Astrophysical Observatory for and on behalf of the National Aeronautics Space Administration under contract NAS8-03060.
This work was partly supported by NASA grant NNX09AC84G, NSF grant AST09-08611, and by the Ministry of Education and Science of the Russian
Federation (contract 11.G34.310001).
Based on observations made with ESO Telescopes at the La Silla Paranal Observatory under programme ID 68.D-0249.\\
This research has made use of SAOImage DS9, developed by
SAO; the SIMBAD and VizieR databases, operated at CDS, Strasbourg, France;
and SAO/NASA's Astrophysics Data System Bibliographic Services.



{\it Facilities:} \facility{CXO (ACIS)}.


\clearpage
\appendix

\section{Reprojection of the X-ray data and X-ray source position}
\label{reproject}
\begin{figure}[b!]
\begin{center}
\includegraphics[width=12.5cm]{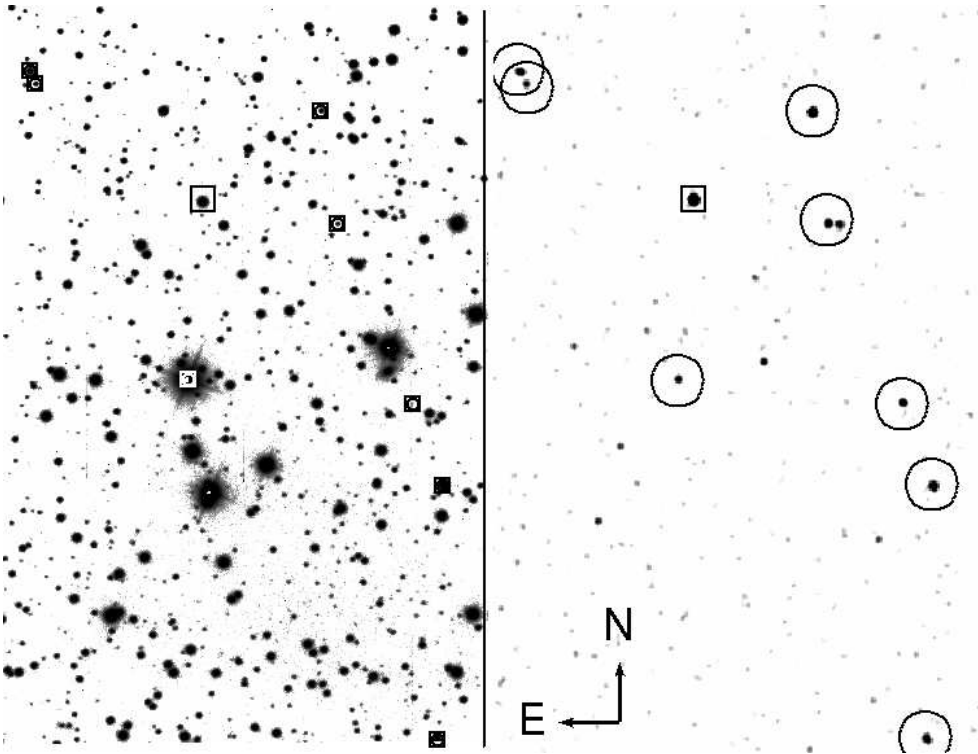}
\caption{Left: the NTT SUSI2 WB655 image is shown for the CCD chip covering the target PSR B1451$-$68. Right: the reprojected X-ray image is plotted for the same coordinate range. The approximate size of each image is $2.5\arcmin \times 3.8\arcmin$. The brightness scale is streched to highlight the respective sources. The X-ray source position of PSR B1451$-$68 is marked by the largest box in the upper part of the images. Smaller boxes in the SUSI2 image indicate optical sources used for the reprojection. The circles in the X-ray image indicate the corresponding X-ray sources. \label{susievli3align}}
\end{center}
\end{figure}

For the optical reference frame, we used the European Southern Observatory (ESO) New Technology Telescope (NTT) observations of PSR B1451$-$68. 
The NTT was equipped with the SUSI2 imager. We chose an image which was obtained with the WB655, a wide $R$-band filter in February 2002.
The field of view of the CCD chip with the target on it is $2.8\arcmin \times 5.5\arcmin$, and there are plenty of 2MASS point sources in this field, e.g., 46 with the highest quality flag \textit{AAA}.
We obtain the absolute astrometry of the ESO SUSI2 image with the help of the 
2MASS point source catalog (PSC, \citealt{Skrutskie2006}) using \textsl{Graphical
Astronomy and Image Analysis Tool}  ($GAIA$) \citep{gaia}. Nominally, an absolute positional accuracy of $3 \sigma_{\rm 2MASS}=300$\,mas \citep{Skrutskie2006} can be achieved for high-quality 2MASS point sources. The rms of the GAIA astrometric fit to the \textit{AAA}-2MASS source positions was determined to be $\sigma_{\rm SUSI2}=178$\,mas. Thus, we expect the source positions in the ESO SUSI2 image to have an \emph{absolute} astrometric $3\sigma$ error of 612\,mas.\\
Nine optical sources in the SUSI2 image were found to correspond to \texttt{wavdetect} X-ray sources on the ACIS S3 chip. Five of them are also 2MASS point sources. We excluded one source close to the expected pulsar position and proceeded with eight optical detections. The sources are indicated in Figure~\ref{susievli3align}.
The X-ray and optical sources were matched and the \textit{Chandra} image astrometry updated by applying the CIAO tasks \texttt{reproject\_aspect} and \texttt{wcs\_match}. 
After removing one poor match, seven sources remained. 
For the seven sources the average residual after reprojecting was $\sigma_{\rm CXO}=130$\,mas in comparison to 340\,mas for the original event file (all without pixel position radomization). 
Thus, our overall \emph{absolute} astrometric uncertainty is $3\, (\sigma^2_{\rm SUSI2}+\sigma^2_{\rm 2MASS}+\sigma^2_{\rm CXO})^{\frac{1}{2}}=3 \sigma^{\rm AM}_{\rm abs}=0.73\arcsec$, while the \emph{relative} astrometric uncertainty of \textit{Chandra} with respect to the 2MASS PSC is $3\,(\sigma^2_{\rm SUSI2}+\sigma^2_{\rm CXO})^{\frac{1}{2}}=3 \sigma^{\rm AM}_{\rm rel}=0.66\arcsec$.

\section{2MASS\,14560002-6843400}
\subsection{Spectral constraints on the nature of 2MASS\,14560002-6843400}
\label{spec2}
Here, we want to test if -- neglecting positional arguments -- 2MASS\,14560002-6843400 could be in principle the counterpart of the X-ray source. 
As described in Sect.~\ref{obs}, we checked different spectral models for correspondence with the energy distribution of the $\sim 200$ source counts within an energy range of 0.3\,keV to 8\,keV; the fit results are listed in Table~\ref{table:fits}.
Amongst others, we checked whether the X-ray spectrum can be described by an optically thin thermal plasma model as one could expect from a stellar corona. 
In Table~\ref{table:fits}, we list the obtained \texttt{apec} model parameter values, the \texttt{mekal} or \texttt{raymond} models gave similar values. 
The \texttt{apec} plasma model fit is acceptable. 
The derived plasma temperature is high, $kT=3.1^{+1.3}_{-0.8}$\,keV. Such high temperatures can be reached in young stars (e.g., \citealt{Preibisch2005,Getman2005}).\\
As a reminder, the derived absorbed X-ray fluxes for the different spectral models range from $2.4 \times 10^{-14}$\,erg\,cm$^2$\,s$^{-1}$ to $2.9 \times 10^{-14}$\,erg\,cm$^2$\,s$^{-1}$ for energies between 0.3\,keV and 5\,keV.
We use the formula by \citet{Maccacaro1988} $\log (f_X/f_V ) = \log (f_X)+0.4 m_V +5.37$ to estimate the optical to X-ray flux ratio.
2MASS\,14560002-6843400 has an apparent $V$-band magnitude of $m_V=16.2$\,mag in the General Guide Star Catalogue version 2.3.2 \citep{Lasker2008}, and $m_V=15.7$\,mag in the NOMAD catalogue \citep{Zach2004}. 
Assuming that the detected X-ray source is the counterpart of 2MASS\,14560002-6843400, the $\log (f_X/f_V )$ range is accordingly from $-1.7$ to $-2.0$. Such flux ratios are typical for K or M stars (e.g., \citealt{agueros2009}). 

2MASS\,14560002-6843400 has the following magnitude measurements:
$J$= $14.50 \pm 0.04$\,mag,
$H$= $13.94 \pm 0.04$\,mag,
$K$= $13.83 \pm 0.06$\,mag,
$B2$= 16.83\,mag (USNO B1),
$R2$= 15.05\,mag (USNO B1),
$I$= 15.11\,mag (USNO B1).
The obtained colors $B2-V$ and $V-I$, are consistent with a late G to early K star, the colors $J-H$, $H-K$, $V-J$, $V-H$, and $V-K$ indicate a K0 to K2 star following intrinsic stellar colors listed by \citet{Currie2010} and \citet{Covey2007}. A K2 dwarf would be consistent with all colors without requiring significant extinction.
From the $N_{\rm H} = 6.5^{+7.5}_{-4.7} \times 10^{20}$\,cm$^{-2}$ obtained for the \texttt{apec} model fit we estimate the optical extinction $A_V=0.36^{+0.44}_{-0.26}$ using the relation between the hydrogen column density and extinction by \citet{Predehl1995}. Using the extincion relations by \citet{Cardelli1989} and $A_V=0.36$ we estimate color corrections and find that a G9 dwarf is the most likely extinction-corrected counterpart for the cataloged NIR and optical magnitudes.
Assuming $M_J=4.2$\,mag according to the Padova tracks \citep{Bertelli2008}, the reported apparent $J$ magnitude translates into a distance of 1.1\,kpc. 
For a K2 dwarf without extinction, assuming $M_J=4.8$\,mag, the reported apparent $J$ magnitude translates into a distance of 0.9\,kpc.

A typical K or late G dwarf main sequence star is expected to have its X-ray peak emission below 1\,keV and show nearly no emission above 1\,keV. Young dwarf stars, on the other hand, could produce the observed X-ray spectrum. There is no obvious star formation region within 1 degree of the target position, and as noted above, 2MASS\,14560002-6843400 has a very slow proper motion ($<12$\,mas\,yr$^{-1}$).
While one cannot exclude an in-situ young stellar object, there is no indication for this  from extinction, outflows, or flares, and a late G or early K dwarf main sequence star seems to be the most likely counterpart of 2MASS\,14560002-6843400. 
From this in turn, we conclude that the contribution of 2MASS\,14560002-6843400 to the X-ray emission is small, and that the X-ray source is not the counterpart of 2MASS\,14560002-6843400, but of the pulsar. \\

\subsection{Investigation of possible source blending}
\label{blending}
\begin{figure*}
\begin{center}
\includegraphics[width=14cm]{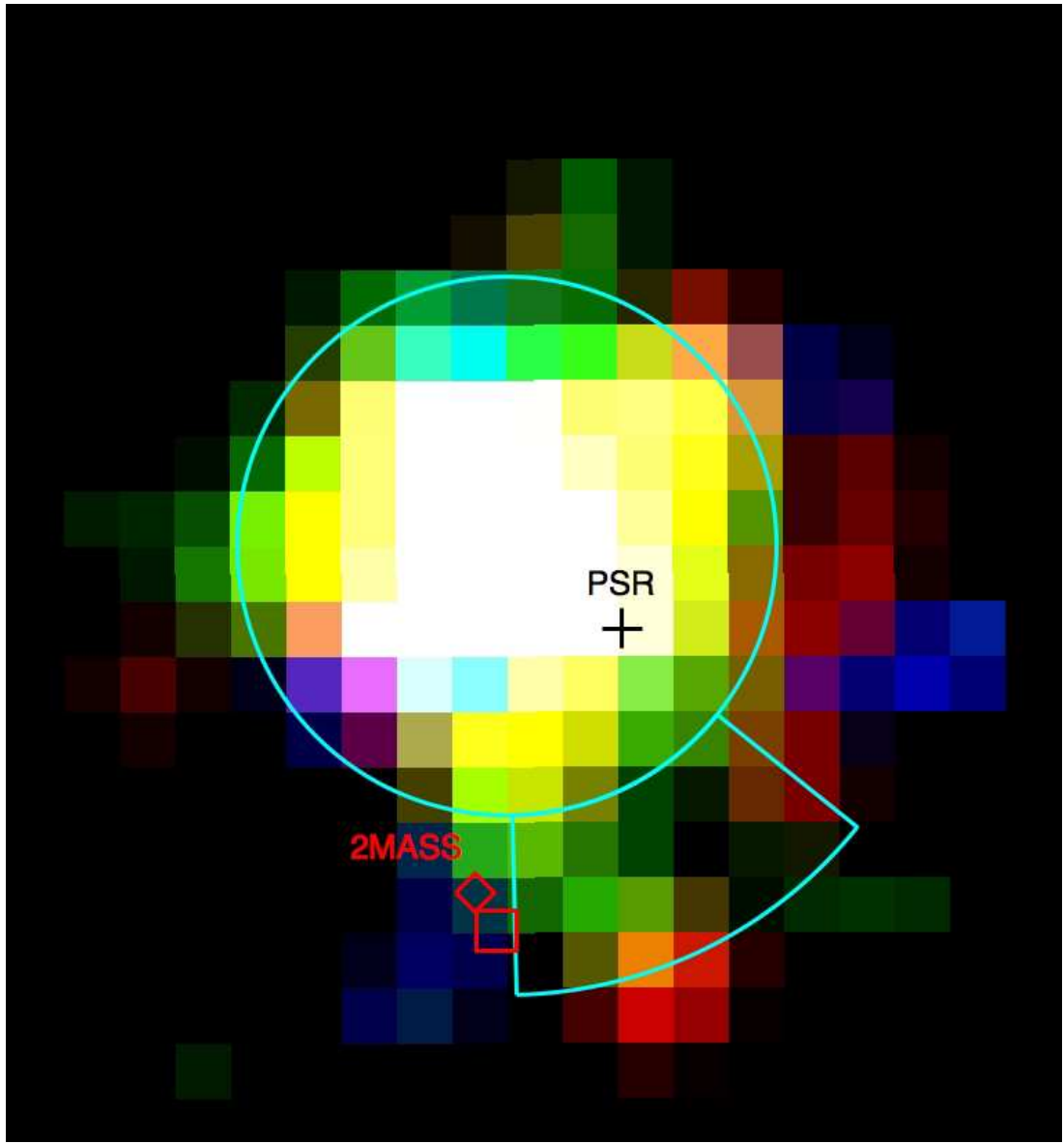}
\caption{This RGB image shows the reprojected event file with the CIAO \texttt{EDSER} subpixel algorithm applied. A 0.25 subpixel binning, as well as slight smoothing (Gauss kernel 2 subpixel), was used in each band. The red band shows the events from 0.3\,keV to 1\,keV, the green band shows the events from 1\,keV to 2\,keV, the blue band shows the events from 2\,keV to 8\,keV. The scale is linear and the same for all bands. North is up, East is to the left. 
The expected pulsar position is marked with a black cross, the 2MASS catalog position of a nearby star is marked by a red diamond, the position of the star in the SUSI2 image is marked with the red square. The big cyan circle has a radius of $0.6\arcsec$ and marks the centroid position of the X-ray source.
The other cyan region indicates the area of the \textsl{Chandra} point spread function asymmetry as inferred by applying the CIAO task \texttt{make\_psf\_asymmetry\_region}.
See text for the discussion.}
\end{center}
\label{subpixrgb}
\end{figure*}

We wish to obtain an upper limit for any potential X-ray flux contribution of the 2MASS star to the X-ray source detected at the pulsar position. 
First, we look at the spatial distribution of the dominant count energies in the X-ray image. Figure~\ref{subpixrgb} shows a color-coded
 image of the reprojected event file with the CIAO \texttt{EDSER} subpixel algorithm \citep{Li2004} applied. Apparently, there is emission near the 2MASS source. However, the location of the 2MASS soure is very close to a region where enhanced emission is expected due to the recently found asymmetry in the \textit{Chandra} point spread function\footnote[4]{http://cxc.harvard.edu/ciao/caveats/psf\_artifact.html} (PSF). For our observation we obtained the inflicted region by applying the CIAO task \texttt{make\_psf\_asymmetry\_region}, it is marked in the figure in cyan and the artifact can constitute $\sim 5$\% of the total brightness. 
If a normal K star at the position of the 2MASS object contributes to the X-ray flux, one would expect an enhancement in soft counts towards this position.
However, there appear to be slightly more hard counts in the direction of the 2MASS source. 
A hard spectrum is unusual for a normal star, less so for a young stellar object. As discussed in Appendix~\ref{spec2}, the 2MASS source is unlikely to be a young stellar object. \\

\begin{figure}
\includegraphics[width=8cm]{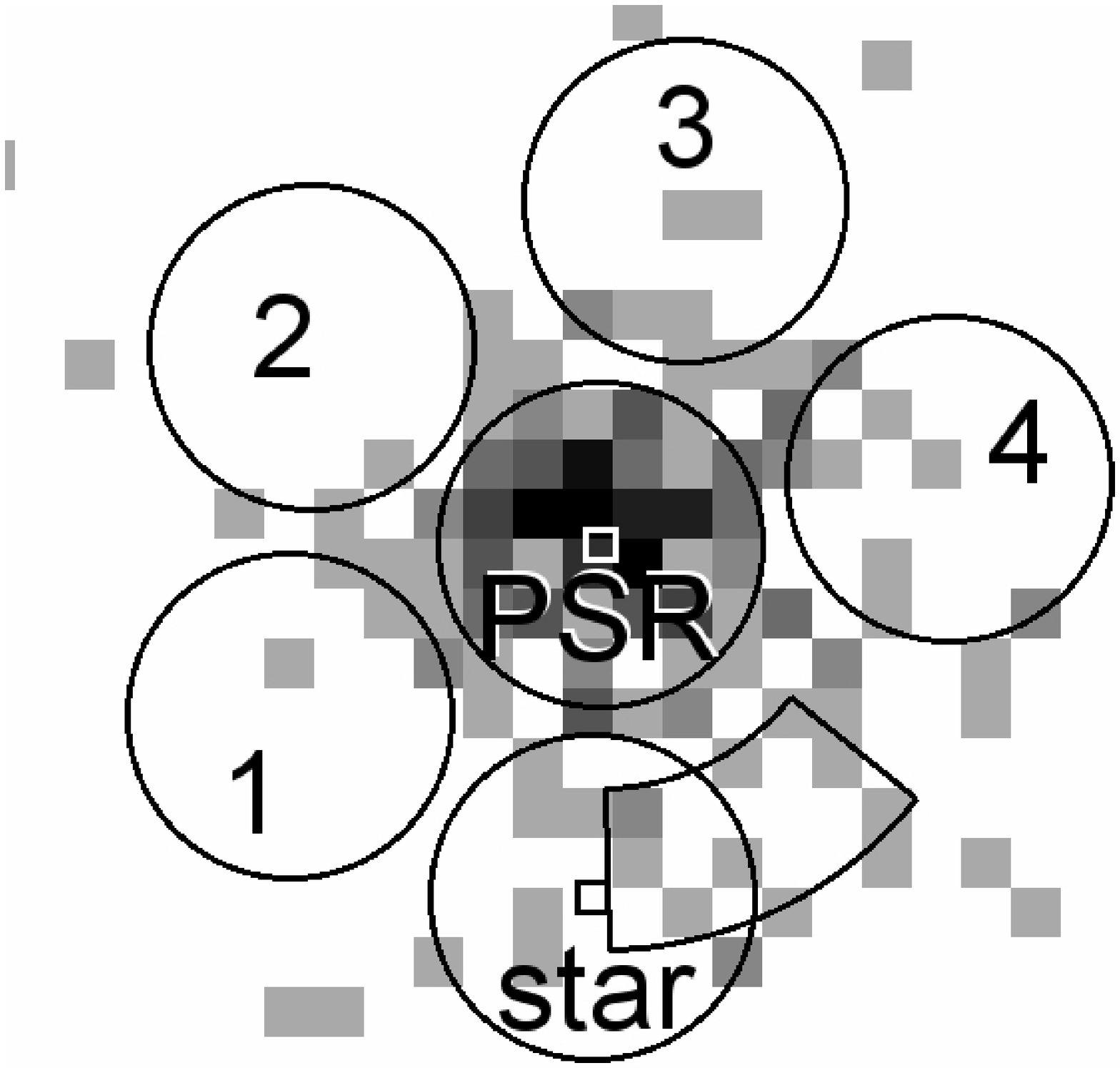}
\caption{This image shows same-size source regions on the reprojected event file with the CIAO \texttt{EDSER} subpixel algorithm applied. The binning is 0.25 \textsl{Chandra} ACIS sky pixel. The `star' circle is centered on the SUSI2 optical position of the 2MASS point source -- its center position is marked by the black box. Similarly, the white box marks the centroid position of the X-ray source. All circles have radii of $0.4\arcsec$. In the southwest, the region of the \textit{Chandra} PSF asymmetry is indicated as obtained with the CIAO task \texttt{make\_psf\_asymmetry\_region}. 
See text for the discussion.}
\label{subpixcircles}
\end{figure}

Next, we measure the number of counts in similar regions around the center of the X-ray source to quantify the overabundance of counts towards the 2MASS star. Figure~\ref{subpixcircles} shows the chosen counting regions: The `star' circle is centered on the 2MASS source's 2002 optical SUSI2 position. Circles 1 to 4 have similar separations from the X-ray centroid and the same size as the `star' region. They are used to obtain the average expected number of counts from the PSF wings for this particular separation and area. This assumes a symmetric PSF (see comments below). To relate the `star' counts, $C_{\rm star}$, to the pulsar counts, $C_{\rm PSR}$, we use a circle of the same size at the position of the X-ray centroid.
We estimate the ratio between the extra `star' counts and the pulsar counts as $(C_{\rm star} - (C_1+C_2+C_3+C_4)\, 0.25) / C_{\rm PSR}$.
Of course, this is a very crude estimation neglecting a detailed \textsl{Chandra} PSF model in general and the outer PSF wings in particular. 
In the ideal case of well separated point sources with similar spectra and using the same aperture for each source, their flux ratios should be constant for different aperture sizes.
In our case we need to account for a flux contribution due to the overlapping of the individual PSFs. This contribution is significant for the potential `star' source, hence the subtraction of $0.25\, (C_1+C_2+C_3+C_4)$  is an approximation of this contribution. The flux contribution of the `star' PSF to $C_{\rm PSR}$ is negligible as seen from \textit{Chandra} MARX simulations.      
For circles with $r=0.4\arcsec$ we infer `star' percentages of 6.3\% for the event file with the \texttt{EDSER} subpixel algorithm applied, and 0\% (actually -1.3\%) for the event file without randomization.
For circles with $r=0.3\arcsec$ we infer `star' percentages of 7.1\% for the event file with the \texttt{EDSER} subpixel algorithm applied, and 3.1\% for the event file without randomization.
Note that the ACIS encircled energy is 50\% for an aperture radius of $r=0.418\arcsec$, while it is 90\% for $r=2\arcsec$ , the radius used in Sections \ref{obs} and \ref{thepulsar} ({\it{Chandra}} Proposers' Observatory Guide, sect. 6.6 \footnote[5]{http://cxc.harvard.edu/proposer/POG/html/index.html}). 
Due to the Poisson noise character of the very few counts we are dealing with, it is difficult to obtain a rigorous error estimate for the above formula. Changing slightly the positions of the individual counting regions, we approximate the error to be around 1\% of the pulsar counts.  
The known PSF assymmetry region of enhanced flux is actually partly overlapping our `star' circle (see Fig.~\ref{subpixcircles}). Since part of the enhanced flux in the `star' circle is likely due to the asymmetry `leakage', our estimated percentages are indeed conservative upper limits. 

\begin{figure}[ht]
\includegraphics[width=15cm]{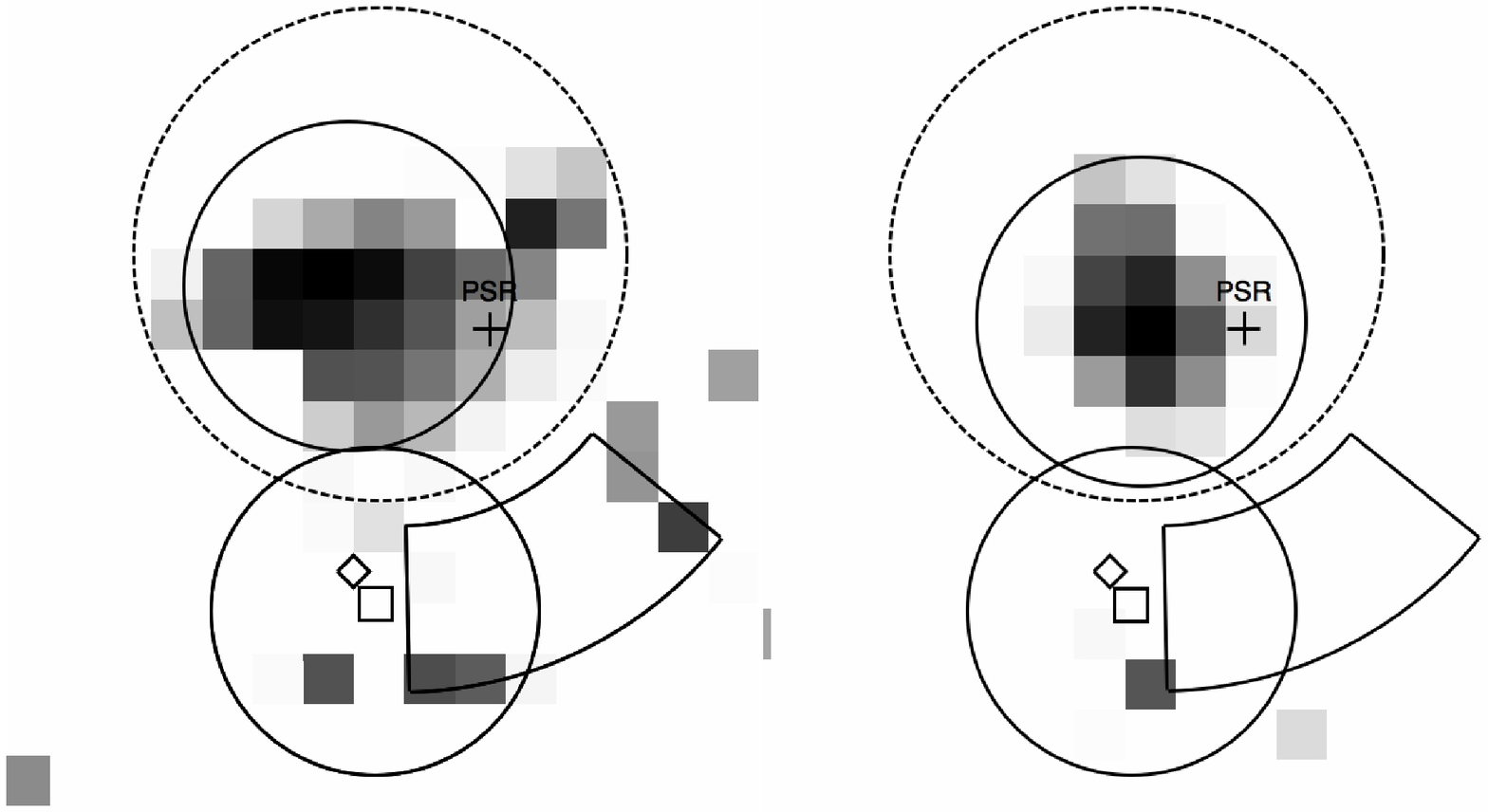}
\caption{On the left is the deconvolved image produced from the event file without randomization, obtained by applying the CIAO task \texttt{arestore}. The right image is the deconvolved image, which was produced from a simulated event file with 2 sources - one with 95\% flux and the other with 5\% flux of the detected X-ray source in our data set. The dashed circle with $r=0.6\arcsec$ is centered at the centroid position of the original event file of our observation. The pulsar position is marked with a cross, the 2MASS point source position with a diamond, the 2002 SUSI2 optical position of this 2MASS source with a small box.
The region of the PSF asymmetry is also indicated.
One of the two solid, $r=0.4\arcsec$, circles in each image is centered on the corresponding centroid position of the deconvolved main source, the other is centered on the SUSI2 position. 
Both images are in log brightness scale to emphasize the low count numbers. 
See text for the discussion.}
\label{arestore}
\end{figure}

We also deconvolved the image applying the \textit{Chandra} Ray Tracer (ChaRT), MARX (v. 4.4 and 4.5) and the CIAO task \texttt{arestore} following standard CIAO threads. Since the subpixel algorithm is not yet implemented in the simulators, we discuss here only the event file without pixel radomization. 
We apply ChaRT \citep{Chart2003} to construct the PSF at the source position. As the input spectrum, we used the parameters derived from the PL fit of the source spectrum (see Sect.~\ref{thepulsar}) in an energy range from 0.3 to 8\,keV. From the centroid source position we obtain the off-axis angle, $\theta=18\arcsec.6$, and the mirror spherical coordinate azimuth, $\phi=327^{\circ}.9$\footnote[6]{For details on the \textit{Chandra} coordinate systrems see http://cxc.harvard.edu/contrib/jcm/ncoords.ps}.
We use a long exposure time (180\,ks) in order to generate a high signal-to-noise-ratio PSF image which is advantageous for the following deconvolution and the source extent estimate.  
The raytrace file is then used within MARX\footnote[7]{http://space/mit/edu/CXC/MARX/index.html} to
produce a PSF image with 0.25 pixel binning.
We set the $DitherBlur$ parameter to $0.2\arcsec$ as recommended in the MARX manual for ACIS data without pixel randomisation.  
Using the PSF image, the CIAO task \texttt{srcextent} reports the X-ray source to be not extended at 90\% confidence.
We then deconvolved the event file applying the task \texttt{arestore} which is based on the Lucy-Richardson deconvolution algorithm \citep{Lucy1974}. The image is shown in Figure~\ref{arestore}.
The new centroid position of the main source in the deconvolved image has a small offset of $\sim 0.1\arcsec$ with respect to the centroid position in the original image. Comparing the count numbers in a $r=0.4\arcsec$ circle centered on the former and a circle of the same size centered on the SUSI2 optical position of the 2MASS star, we derive a flux percentage of 6.5\% for the counts around the star with respect to those of the pulsar.
Using the original centroid position for the main source decreases this percentage to 6.1\%.
The deconvolution produces `extra' counts not only towards the direction of the 2MASS source, which is located at one edge of the PSF asymmetry region, but also southwest of the pulsar where the other edge of the PSF asymmetry region lies. Since neither ChaRT, MARX, nor \texttt{arestore} account for this asymmetry, some artefacts in the south-southwest direction can be expected after deconvolution.\\
For comparison, we also used MARX to simulate two sources having combined as many counts as the $r=2\arcsec$ main source region -- one source with 95\% of the flux at the main source centroid position, and one with 5\% flux at the SUSI2 optical position of the 2MASS star. Again, we set the  $DitherBlur$ parameter to $0.2\arcsec$ and proceeded with the processing as described above. Again we find an $\sim 0.1\arcsec$ shift (but in another direction) in the centroid position of the deconvolved image comparing to the original input coordinates. While these shifts are all within the nominal \textit{Chandra} astrometric accuracy, we do not utilize the \texttt{arestore} positions for further astrometric purposes. Using again the $r=0.4\arcsec$ circles as indicated in Figure~\ref{arestore}, we find 4.0\% of the simulated main source counts in the circle around the simulated 2MASS source. From this we infer that, in principle, it is possible to recover a $\sim 5$\% X-ray source at a 0.77\arcsec\ separation, which the 2MASS source has from the main X-ray source centroid.\\


\end{document}